\documentclass[a4paper,twoside,american,english,twocolumn]{svjour3}
\usepackage[T1]{fontenc}
\usepackage[latin9]{inputenc}
\usepackage{float}
\usepackage{units}
\usepackage{url}
\usepackage{amsmath}
\usepackage{amssymb}
\usepackage{graphicx}
\usepackage{esint}

\makeatletter

\newcommand{\noun}[1]{\textsc{#1}}

\RequirePackage{fix-cm}

\smartqed  % flush right qed marks, e.g. at end of proof

\makeatother

\usepackage{babel}
\begin{document}

%\selectlanguage{american}%

\title{Properties of gravity near the Schwarzschild radius and the cosmological
redshift}

\titlerunning{Properties of gravity near the Schwarzschild radius\foreignlanguage{english}{}}

%\selectlanguage{english}%

\author{Leonid V. Verozub }

%\selectlanguage{american}%

\authorrunning{Properties of gravity near the Schwarzschild radius\foreignlanguage{english}{}}

%\selectlanguage{english}%

\institute{L. Verozub \at  Kharkov Karasin University, Kharkov 61077, Ukraine\\
\email{leonid.v.verozub@univer.kharkov.ua}\\
}

\date{Received: date / Accepted: date}
\maketitle
%\selectlanguage{american}%
\begin{abstract}
The radius of the observable region of the Universe is of the order
of its Schwarzschild radius. Due to the spherical symmetry, this allows
to check the properties of the gravitational force in the vicinity
of the Schwarzschild radius by comparing the theoretical and observed
Hubble diagram at high redshifts. This can be done in a simple model
that follows from projective-invariant equations of gravitation.This
paper shows that the Hubble diagram up to $z=8$ testifies in favor
of the specific properties of gravity near and inside of the Schwarzschild
radius.\foreignlanguage{english}{. }

%\selectlanguage{english}%
\end{abstract}
%\selectlanguage{english}%

\section{Introduction\label{sec:Introduction}}

%\selectlanguage{american}%
Properties of gravity near the Schwarzshild radius are key test of
general relativity. This paper is based on the fact that the radius
of the observable Universe is of the order of its Schwarzschild radius.
Starting from this fact, we show that observed properties of the Hubble
diagram at high redshifts testify in favor of specific properties
of gravity, which follow from the projective invariant equations of
gravitation \cite{Verozub2008a}. This confirms results of paper \cite{Verozub2006},
where it was shown that such properties of gravity do not contradict
observations near the supermassive object at the center of our galaxy.

In paper \cite{Verozub2008a} has been shown that the properties of
gravitation may differ significantly from those resulting from general
relativity near the Schwarzschild radius and less than that. There
is no an event horizon. The force acting on a freely falling test
particle to a dot mass becomes repulsive near the Schwarzschild radius
$r_{g}$, and tends to zero when the distance to the center tends
to zero.

The lack of the event horizon and the weakness of the gravitational
force near the center makes it possible the existence of super-massive
objects without an event horizon. Such objects are candidates to the
supermassive objects that exist in centers of galaxies \cite{Verozub2006,Verozub2008b}.

The properties of gravity near the Schwarzschild radius should be
manifested in the magnitude of the velocity of galaxies at very high
redshifts, and therefore, in the Hubble diagram. It allows to test
gravity near the Schwarzschild radius. Fortunately, in addition to
the great data from observations of supernovae Ia \cite{Riess} ,
we have now the numerous data obtained from observations of gamma-ray
bursts up to $z=8$ \cite{Cardone,Schaefer,Demiansky}. The mentioned
data are not direct observations. They depend in part on the used
Lambda CDM model . Therefore, their accuracy is not high. However,
they can make a qualitative conclusion about the properties of gravity
near the Schwarzschild radius.

\section{Meaning of metric-field approach to theory of gravitation\foreignlanguage{english}{\label{sec:2}}}

\subsection{Space-time geometry and reference system\foreignlanguage{english}{\label{sub:2.1}}}

In the early twentieth century, Henri Poincaré realized that geometry
of space and time have no physical meaning by itself, without the
knowledge of properties of the measuring instrument. Only the combination
\textquotedbl{}geometry + measuring instruments\textquotedbl{} has
a checked on experience meaning \cite{Verozub2008a,Verozub2013}.
This applies also to the properties of space-time. Strictly speaking,
we cannot speak about the properties of space-time without knowledge
of properties of a reference frame, and we cannot speak about properties
of the reference frame without knowledge of properties of space-time,
because the reference frame is an instrument for investigation of
space-time properties. Each of these concepts has no physical meaning
in itself.

Based on the assumption that space-time in the inertial reference
frames is pseudo-Euclidean, and starting from the conception of relativity
of space-time, we can find a space-time metric for a certain class
of non-inertial reference frames.

%\selectlanguage{english}%
We believe that the concept of \textquotedbl{}frame of reference\textquotedbl{}
has a physical meaning only when it is defined by some operational
manner. The same can be said about a \textquotedbl{}comoving\textquotedbl{}
coordinate system. This coordinate system has a physical meaning only
if it is obtained by some transformation of the orthogonal coordinate
system of pseudo-Euclidean space-time. Therefore, we define the non-inertial
reference body frame as a system of material points which move under
the influence of a force field defined in the inertial frame in Minkowski
space-time.

%\selectlanguage{american}%
Thus, consider \cite{Verozub2008a,Verozub2011} a non-inertial frame
of reference $B$, reference body of which is formed by identical
dot masses $m$, moving under action of some force field $F(x)$ given
in an inertial reference system (IRF) $A$, where space-time is pseudo-Euclidean.
The reference frame $B$ will be here called the proper reference
frame (PRF) of the force field $F(x)$. If an observer, located in
$B$, rejects the Newtonian idea of absolute space and time, and believes
that space-time is relative in the sense of Berkeley-Leibniz-Mach,
then from his point of view the above dot masses are the points of
his physical space. From his point of view they are at rest both in
the non-relativistic and the relativistic sense. Consequently, their
world lines are geodesics of space-time in the frame $B$ so that
the equality $\delta\int ds=0$ holds, where $ds$ is the line element
of space-time in $B$.

However, on the other hand, the motion of the material points in the
frame $A$ (in Minkowski space-time) is described by a Lagrangian
action $S=\int L(x,\dot{x})dt$. It follows from this fact that the
line element of space-time in the frame $B$ is of the form 
\begin{equation}
ds=k\, dS,\label{eq:dsmy-1}
\end{equation}
where $k$ is a constant and $dS=L(x,\dot{x})dt$.

In the limit $F(x)\rightarrow0$ 
\begin{equation}
S=-mc\int(1-v^{2}/c^{2})^{1/2}dt,
\end{equation}
where $v$ is 3-velocity of particles of the reference body, and $c$
is the speed of light. It follows from this fact that the constant
$k$ should be equal to $-(mc)^{-1}.$

Consider, for example, the following PRF.

The reference body consists of noninteracting electric charges in
an electromagnetic field. In a Cartesian coordinate system the action
describing the motion of the particles can be written as follows \cite{Landau}:
\begin{equation}
S=\int\left(-mc^{2}(1-v^{2}/c^{2})^{1/2}-\frac{e}{c}\, A_{\alpha}(x)\, dx^{\alpha}/dt\right)dt,\label{LagrangGarge}
\end{equation}
where $A_{\alpha}$ is the 4-potential, $e$ is charge of the particles
\footnote{As usual, in this paper Greek letters run from 0 to 3, and Latin -
from 1 to 3.%
}.

For the given reference frame 
\begin{equation}
ds=d\sigma-\frac{e}{mc}\, A_{\alpha}dx^{\alpha},\label{dsInRotationNIFR}
\end{equation}
where $d\sigma$ is the line element of space-time in the IRF. It
is a Finslerian metric.

Of course, such a frame of reference is not similar to the accelerated
reference frame formed by neutral particles. However, this does not
prevent to its theoretical analysis, assuming that the reference body
is formed by identical ions. They can be regarded as an atomic clocks
that are almost unaffected by accelerations.

Based on the fact that the clocks measure the length of its own world
line, one can find the time interval in such PRFs. Evidently, 
\[
dT=\overset{0}{dT}-\frac{e}{mc}\, A_{\alpha}dx^{\alpha},
\]
where $dT=ds/c$ and $d\overset{0}{T}=d\sigma/c$ are proper time
intervals in the PRF and IRF, correspondingly.

The following two cases are of interest:

1. The reference body consists of noninteracting electric charges
in a constant homogeneous electric field $E$ directed along the axis
$x$. According to (\ref{dsInRotationNIFR}) 
\begin{equation}
dT/dt=1-\frac{e}{mc^{2}}\varphi=1+\frac{e}{mc^{2}}Ex,
\end{equation}
where $\varphi=A_{0}$, and $E$ is the electric field strength. Because
the electric force $eE=maw$, where $w$ is the acceleration with
respect to the IFR, this result is equivalent to the well known one:

\begin{equation}
dT/dt=1+\frac{wx}{c^{2}}.
\end{equation}

This result shows that difference between clock in IRF and PRF is
not a kinematic effect, and is caused by a force field.

The same result is obtained for the PRF of the homogeneous field of
the Earth, reference body of which is formed by particles free falling
in the field. The reason is that the replacement $e\varphi$ by the
gravitational potential leads to the same equation of the motion of
test particles as the equations for charges.

2. The reference body consists of noninteracting electric charges
in a constant homogeneous magnetic field $H$ directed along the axis
$z$.

It follows from the Stokes theorem that in this case the modulus $A$
of the potential $A_{i}$ of a particle at the distance $r$ from
the center of the orbit of the reference body is equal $A=Hr/2$,
which shows that $A$ is a modulus of the 3-vector $\vec{A}=\tfrac{1}{2}\vec{B}\times\vec{r}$,
and the $\vec{A}$ is directed tangentially to the orbit circle. For
this reason, according to (\ref{dsInRotationNIFR}), 
\begin{equation}
dT/dt=1-\frac{e}{mc^{3}}\frac{Hr^{2}\omega}{2}=1-\frac{F_{l}r}{2\, mc^{2}},
\end{equation}
where $\omega=d\varphi/dt$ is the angular velocity of the body reference,
and $F_{l}=eHr\omega/c$ is the Lorentz force. Since the centrifugal
force $F_{c}=F_{l}=mw_{c}$, where $w_{c}$ is the centrifugal acceleration,
we arrive at the conclusion that this result is equivalent to the
well known one:

\begin{equation}
dT/dt=1-\frac{\omega^{2}r^{2}}{2c^{2}}.
\end{equation}
This result coincides with the result for the rotating disk due to
the fact that the motion of points on the disk can be described by
a similar Lagrangian.

Let us consider another important example.

In papers \cite{VerozubJMP,Verozub2013} has been proved that streamlines
of any perfect isentropic relativistic fluid are geodesic lines in
a Riemannian space-time.

In more detail, the motion of macroscopically small elements (``particles'')
of the fluid can be considered in two ways.

In Minkowski space-time, where $d\sigma^{2}=\eta_{\alpha\beta}\, dx^{\alpha}dx^{\beta}$
is the line element and $\eta_{\alpha\beta}(x)$ is the metric tensor,
this motion can be described by the following Lagrangian 
\begin{equation}
L=-mc^{2}\left(G_{\alpha\beta}\frac{dx^{\alpha}}{dt}\frac{dx^{\beta}}{dt}\right)^{1/2}dt,\label{eq:geolagrang}
\end{equation}
where $G_{\alpha\beta}=\chi^{2}\eta_{\alpha\beta}$, 
\[
\chi=\frac{\vartheta}{nmc^{2}}=1+\frac{\varepsilon}{mc^{2}}+\frac{P}{\rho c^{2}},
\]
$\vartheta$ is the enthalpy per unit volume, $\varepsilon$ is the
fluid density energy, $m$ is the mass of the fluid ``particle'',
and $c$ is the speed of light.

Equations of the motion of the fluid element that arise from this
Lagrangian are the standard equations of the field velocities of an
relativistic isentropic fluid.

On the other hand, in a co-moving reference frame this motion can
be described (in the same coordinate system) as the motion along the
geodesics of the Riemannian space, the line element of which has the
form 
\[
ds^{2}=G_{\alpha\beta}dx^{\alpha}dx^{\beta}.
\]
Indeed, in this case, for an observer located in this frame time is
measured by the length of its world line. When using $s$ as a parameter
of the length it is easy to see that the Lagrange equations give the
standard equations of a geodesic in the Riemannian space-time with
the metric tensor $G_{\alpha\beta}$.

Thus, there are reasons to believe that in PRFs a force field manifests
itself as a curvature of space-time. More precisely, a field $F(x)$
can be considered in two ways: a) as a force field in an IRF in the
Minkowski space-time , and b) as a curvature of space-time in PRFs
according to (\ref{eq:dsmy-1}).

It is difficult to verify our conclusion directly, but in the case
of the gravitational field this conclusion leads to observable physical
consequences since space-time is a bi-metric, and the existence of
a flat metric should have an impact on the field equations. In addition,
we live in the expanding Universe, and so we have the opportunity
to study the phenomenon in a PRF of the gravitational field produced
by matter of the Universe.

From the above point of view, gravity can be considered as a true
field in the Minkowski space-time where the Lagrangian, describing
the motion of test particles in this field, has the form 
\[
L=-mc\,[g_{\alpha\beta}(x)\dot{x}^{\alpha}\dot{x}^{\beta}]^{1/2},
\]
where $g_{\alpha\beta}(x)$ is a tensor field in the the Minkowski
space-time. %
\footnote{This field can in principle be formed by another function $\psi(x)$,
which is some traditional characteristic of field in the Minkowski
space-time \cite{Thirring}.%
}

In this case, according to (\ref{eq:dsmy-1}) the line element of
space-time in PRFs is given by 
\[
ds^{2}=g_{\alpha\beta}(x)dx^{\alpha}dx^{\beta},
\]
that is, space-time in PRFs of gravitational field is a Riemannian
with non-zero curvature, where $g_{\alpha\beta}$ is the metric tensor.

Thus, we suppose that gravity can be considered as a field in an inertial
reference frame in Minkowski space-time, and a space-time curvature
in the proper reference frames.

\subsection{Projective invariance\label{sub:2.2}}

An other starting point of the theory is based on the observation
that the equations of the motion of test particles ( geodesic lines)
are invariant under some group of the continuous transformations -
geodesic (projective) mappings of the Riemannian space-time \cite{Eisenhart},
in any fixed coordinate system.

A diffeomorphism between two pseudo-Riemannian spaces $V_{n}$ and
$\bar{V}_{n}$, with a metric tensor $g$ and $\bar{g}$, respectively,
is called geodesic if it is geodesic-preserving, that is, when it
maps any geodesic of $V_{n}$ into an geodesic of $\bar{V}_{n}$ again.

A necessary and sufficient condition for existence of a geodesic mapping
between $V_{n}$ and $\bar{V}_{n}$ is that the equations

\begin{equation}
\bar{\Gamma}_{\beta\gamma}^{\alpha}(x)=\Gamma_{\beta\gamma}^{\alpha}(x)+\delta_{\beta}^{\alpha}\phi(x)_{\gamma}+\delta_{\gamma}^{\alpha}\phi(x)_{\beta}\label{gaugeTrans}
\end{equation}
are satisfied, where $\Gamma_{\beta\gamma}^{\alpha}(x)$ and $\bar{\Gamma}_{\beta\gamma}^{\alpha}(x)$
are components of the Christoffel symbols in $V_{n}$ and $\bar{V}_{n}$,
respectively.

The condition (\ref{gaugeTrans}) is equivalent of the following Levi-Civita
equations

\[
\bar{g}_{\alpha\beta;\gamma}=2\phi_{\gamma}\bar{g}_{\alpha\beta}+\phi_{\alpha}\bar{g}_{\beta\gamma}+\phi_{\beta}\bar{g}_{\alpha\gamma},
\]
where the semicolon denotes a covariant derivative in $V_{n}$, $\phi_{\alpha}$
is some gradient-like vector, i.e. $\phi_{\alpha}=\partial\phi/\partial x^{\alpha}$
.

For example, if $\Gamma_{\beta\gamma}^{\alpha}$ are Christoffel symbols,
then under using time $t=dx^{0}/c$ as a parameter, the differential
equations of a geodesic line are of the form

\begin{equation}
\overset{..}{x}^{\alpha}+\left(\Gamma_{\beta\gamma}^{\alpha}-c^{-1}\Gamma_{\beta\gamma}^{0}\overset{.}{x}^{\alpha}\right)\overset{.}{x}^{\beta}\overset{.}{x}^{\gamma}=0\label{eq:geodeq}
\end{equation}
where $\dot{x}^{\alpha}=dx^{\alpha}/dt$, $\ddot{x}=d\dot{x}/dt$.
It easily to verify that these equations are invariant under the mapping
(\ref{gaugeTrans}) of the Christoffel symbols in any coordinate system.

Since $\phi_{\alpha}$ is a gradient function, (\ref{gaugeTrans})
are very similar to a generalization of the gauge transformations
of 4-potentials in the classical electrodynamics.

In a supplement we give two example of this fact. Namely, it is shown
that in any coordinate system the FWR and the Schwarzschild metrics
have a continuous set of geodesically equivalent metrics.

It is obviously, all the connection coefficients, which are connected
by such transformation, are physically equivalent. They describe the
same gravitational field, because a classical field is defined by
properties of motion of test particles. (Just as the 4-potentials
in classical electrodynamics, connected by a gauge transformation
$A_{\beta}\rightarrow A_{\beta}+\partial_{\beta}\phi(x)$).

The projective transformations of the connection coefficients induce
transformation of the metric tensor, the curvature tensor and the
Ricci tensor. This is a reason that the classical Einstein's equations
are not invariant under projective mappings of Riemannian spaces \cite{Petrov},\cite{Mikes}
which obviously should play a role of gauge transformations.

It follows from (\ref{gaugeTrans}) that $\bar{\Gamma}_{00}^{i}=\Gamma_{00}^{i}$.
This is a reason why geodesic invariance is not manifest itself in
non-relativistic theory. It is important only in relativistic theory
of gravitation.

Simple geodesic-invariant generalization of Einstein's equations has
examined in \cite{Verozub2008a} . These equations are bi-metric,
i.e. these equations contain the Minkowski metric tensor $\eta_{\alpha\beta}(x)$,
and $g_{\alpha\beta}(x)$. In cosmology this means that we have two
possibilities. Either we are considering expanding Universe locally
from the viewpoint of an observer in an inertial reference frame where
space-time is supposed to be Minkowskian, either from the viewpoint
of an observer in a co-moving frame, in which space-time is Riemannian.
Both possibilities are locally in principle equivalent, if they are
described by some appropriate differential equations for finding the
functions $g_{\alpha\beta}(x)$.

The fact that the used usually geometrical characteristics of space-time
are not projective-invariant (like 4-potentials in electrodynamics
are also not gradient-invariant) means that they cannot be considered
as observable characteristics of gravitational field. The problem
is to find the geometric objects which are invariant under geodesic
mappings of Riemannian spaces. Two such objects are well known. These
are the Weyl tensor and also the Thomas symbols \cite{Eisenhart},
which are some natural geodesically invariant generalization of the
Christoffel's symbols.

A geodesic-invariant generalization of the metric tensor can also
be defined \cite{VerKoch2000}. Such object is based on a 5-dimensional
interpretation of projective mappings in homogeneous coordinates,
and is of the form

\begin{equation}
\hat{g}(x)_{\alpha\beta}=g_{\alpha\beta}(x)-f_{\alpha}(x)f_{\beta}(x).\label{eq:gikinvar}
\end{equation}
In this equation 
\[
f_{\alpha}(x)=\frac{1}{2}\frac{\partial}{\partial x^{\alpha}}\ln\left(\frac{g}{\eta}\right),
\]
where $g$ and $\eta$ are the determinants of space-time in the PRF
and flat metric, correspondingly. However, in this case, geodesic
mappings $\overline{g}_{\alpha\beta}(x)\rightarrow g_{\alpha\beta}(x)$
are not arbitrary.

\section{Spherically symmetric solution in infinite medium\label{sec:3}}

Based on the discussion above in section 2 we consider a simple relativistic
model of the homogeneous and isotropic Universe as a self-gravitating
expanding dust-like matter, space-time of which is a Riemannian in
co-moving reference frame, and is a Pseudo-Euclidean in inertial reference
frame.

The simplest geodesic-invariant generalizations of the vacuum Einstein
equations are \cite{Verozub2008a}: 
\begin{equation}
B_{\alpha\beta;\gamma}^{\gamma}-B_{\alpha\delta}^{\epsilon}B_{\beta\epsilon}^{\delta}=0.\label{myeqs}
\end{equation}
These equations are bi-metric differential equations for the tensor
\begin{equation}
B_{\alpha\beta}^{\gamma}=\Pi_{\alpha\beta}^{\gamma}-\overset{\circ}{\Pi}_{\alpha\beta}^{\gamma},\label{tensB}
\end{equation}
where $\Pi_{\alpha\beta}^{\gamma}$ and $\overset{\circ}{\Pi}_{\alpha\beta}^{\gamma}$
are the Thomas symbols of Riemannian and Minkowski space-time, respectively.
They are given by equations 
\begin{equation}
\Pi_{\alpha\beta}^{\gamma}=\Gamma_{\alpha\beta}^{\gamma}-(n+1)^{-1}\left[\delta_{\alpha}^{\gamma}\Gamma_{\epsilon\beta}^{\epsilon}+\delta_{\beta}^{\gamma}\Gamma_{\epsilon\alpha}^{\epsilon}\right],\label{Thomases}
\end{equation}

\begin{equation}
\overset{\circ}{\Pi}_{\alpha\beta}^{\gamma}=\overset{\circ}{}{\Gamma}_{\alpha\beta}^{\gamma}-(n+1)^{-1}\left[\delta_{\alpha}^{\gamma}\overset{\circ}{}{\Gamma}_{\epsilon\beta}^{\epsilon}+\delta_{\beta}^{\gamma}\overset{\circ}{}{\Gamma}_{\epsilon\alpha}^{\epsilon}\right],\label{Thomases0}
\end{equation}
$\overset{\circ}{\Gamma}_{\alpha\beta}^{\gamma}$ are the Christoffel
symbols of the Minkowski space-time , $\Gamma_{\alpha\beta}^{\gamma}$
are the Christoffel symbols of the Riemannian space-time whose fundamental
tensor is $g_{\alpha\beta}$. A semi-colon denotes a covariant differentiation
in flat space-time.

The Lagrangian $L$ describing the motion of test particles in a static
spherically symmetric homogeneous and isotropic dust-like medium ,
which is invariant under the mapping $t\rightarrow-t$, reads:

\begin{equation}
L=-mc[A\dot{r}^{2}+B(\dot{\theta}^{2}+\sin^{2}\theta\ \dot{\varphi}^{2})-c^{2}C]^{1/2}\label{LagrTestParticls}
\end{equation}
where $A$ ,$D$ and $C$ are functions of the radial coordinate $r$.

The associated line element of the space-time in PRFs is given by

\begin{align}
ds^{2} & =A\, dr^{2}+B[\, d\theta^{2}+\sin^{2}\theta\, d\varphi^{2}]-C\,(dx^{0})^{2},\label{eq:ds-2}
\end{align}
The functions $A(r),\, B(r),\, C(r)$ should be found from the field
equations (\ref{myeqs}) . Because of the projective invariance of
the gravitation equations, some additional (gauge) conditions can
be imposed on the Christoffel symbols. In particular, at the conditions

\begin{equation}
Q_{\alpha}=\Gamma_{\alpha\sigma}^{\sigma}-\overset{\circ}{\Gamma}_{\alpha\sigma}^{\sigma}=0
\end{equation}
the gravitation equations are reduced to Einstein's vacuum equations
$R_{\alpha\beta}=0$. Therefore, the spherically-symmetric solution
$A(r)$, $B(r)$, $C(r)$ can be found as a solution of the system
of the differential equations :

\begin{equation}
R_{\alpha\beta}=0\label{eq:ricci}
\end{equation}
and 
\begin{equation}
Q_{\alpha}=0,\label{eq:dopcond}
\end{equation}

It must be stressed that the condition (\ref{eq:dopcond} ) is a tensor
equation. It does not impose any conditions on the coordinate system.

Thus the classical Einstein equations are eqs. (\ref{myeqs}) at a
specific gauge condition.

Let us find solutions of (\ref{eq:ricci},\ref{eq:dopcond}) \foreignlanguage{english}{which
satisfy the conditions at infinity: 
\begin{equation}
\lim\limits _{r\rightarrow\infty}A(r)=1,\;\lim\limits _{r\rightarrow\infty}(B(r)/r^{2})=1,\;\lim\limits _{r\rightarrow\infty}C(r)=1.\label{limitConditions}
\end{equation}
}

%\selectlanguage{english}%
Usually such conditions at infinity are meaningful only for the spatially
restricted distribution of matter. However, due to the spherical distribution
of matter and specific properties of gravity, which are derived from
the projective invariant equations, they are valid for any infinite
homogeneous material medium.

Conditions (\ref{eq:dopcond}) yield one equation: 
\begin{equation}
B^{2}AC=r^{4}.
\end{equation}
It allows to exclude the function $A$ from the (\ref{eq:ds-2}).
Then the equations $R_{11}=0$ and $R_{00}=0$ take the form: 
\begin{equation}
-2BC^{\prime}+2rB^{\prime}C^{\prime}+rBC^{\prime\prime}=0,\label{EinstEq1}
\end{equation}
\begin{multline}
-4BCB^{\prime}+rCB^{\prime2}-2BC^{\prime}+2rBB^{\prime}C^{\prime}+2rBCB^{\prime\prime}\\
+rB^{2}C''=0\label{eq:EinstEq2}
\end{multline}

%\selectlanguage{american}%
Because of equality ( \ref{EinstEq1}) the sum of tree terms in (\ref{eq:EinstEq2})
is equal to zero, and we obtain a differential equation for the function
$B(r)$: 
\begin{equation}
2rBB^{\prime\prime}+rB^{\prime2}-4BB^{\prime}=0.
\end{equation}
A general solution of this equation can be written as $B=a(r^{3}+\mathcal{K}^{3})^{2/3}$
where $a$ and $\mathcal{K}$ are some constants.

Now the function $C$ can be found from differential equation (\ref{EinstEq1})
in the form 
\begin{equation}
C^{\prime\prime}+2\frac{rB^{\prime}-B}{rB}C^{\prime}=0,
\end{equation}
where $(rB^{\prime}-B)/rB=(r^{3}-\mathcal{K}^{3})/(r^{4}+r\mathcal{K}^{3})$.
A general solution of this equation is $C=b-\mathcal{Q}/f$ where
$f=(r^{3}+\mathcal{K}^{3})^{1/3}$, and $\mathcal{Q}$ are a constant.
It follows from the Newtonian limit that $b=1$ and $\mathcal{Q}=2GM/c^{2}$
is the Schwarzschild radius of mass $M$.

Therefore, the solution for spherically-symmetric field is given by:

\begin{eqnarray}
C & = & 1-r_{g}/f,f=(r^{3}+r_{g}^{3})^{1/3},B=f^{2},A=f'^{2},\label{eq:ABC}
\end{eqnarray}
\foreignlanguage{english}{where 
\begin{equation}
f=(r^{3}+\mathcal{K}^{3})^{1/3},f'=df/dr.
\end{equation}
}

%\selectlanguage{english}%
The constant $\mathcal{K}$ cannot be obtained from (\ref{limitConditions})
or from the consideration of non-relativistic limit, However, one
can use a physical argumentation. Namely, we demand that the spherically
symmetric solution must not have a singularity at the center.

%\selectlanguage{american}%
It is enough to set $\mathcal{K=Q=}r_{g}$ . In this case, in the
coordinate system used the line element of space-time is of the form:

\begin{equation}
ds^{2}=-\frac{f^{\prime2}dr^{2}}{(1-r_{g}/f)}-f^{2}[d\theta^{2}+\sin^{2}\theta\; d\varphi^{2}]+(1-\frac{r_{g}}{f})dx^{0}{}^{2},\label{MyMetricFormOfSpaceTime}
\end{equation}
where $f=($ $r_{g}^{3}$ $+$ $r^{3})^{1/3}$. This solution has
no the event horizon and no singularity in the center.

Tyis equation formally coincides with the original Schwarzschild solution
\cite{Abrams 1} of the Einstein equations. It is a particular solution
of the equations at the condition $det|g_{\alpha\beta}(x)|=1$:

We suppose that the lack of the singularity is sufficiently important
reason to investigate just this case. Especially because it is not
contradict the available observations \cite{VerKoch2000,Verozub2006}.
For this reason, the solution (\ref{eq:ABC}) with $\mathit{\mathcal{K}=r_{g}}$
and with our conditions at infinity will be a basis for further consideration.

At the above conditions the equations of the motion of test particles
in the plane $\theta=\pi$/2 can be obtained by the law of conservation
of the energy $E$ and the angular moment $J$:

%\begin{singlespace}

\begin{center}
$\dot{r}\,\partial L/\partial\dot{r}-\dot{\varphi}\partial L/\partial\dot{\varphi}-L=E$ 
\par\end{center}

%\end{singlespace}

and

\[
\partial L/\partial\dot{\varphi=J,}
\]
where $\dot{r}=dr/dt$ $\dot{\varphi}=d\varphi/dt$.

These equations are of the form

\begin{align}
\overset{.}{r}^{2} & =c^{2}C^{2}\left[1-\varkappa^{2}C/\overline{E}^{2}(1+\overline{J}^{2}/\varkappa^{2}\overline{r}^{2})\right]\label{eq:rprime}\\
\overset{.}{\varphi} & =cC\overline{J}r_{g}/\overline{r}^{2}\overset{\_}{E}\notag\label{eq:fiprime}
\end{align}
where $\overline{E}=E/mc,$ $\overline{J}=J/r_{g}mc$.

\begin{figure}[H]
\begin{minipage}[t]{0.45\columnwidth}%
\includegraphics[width=4cm]{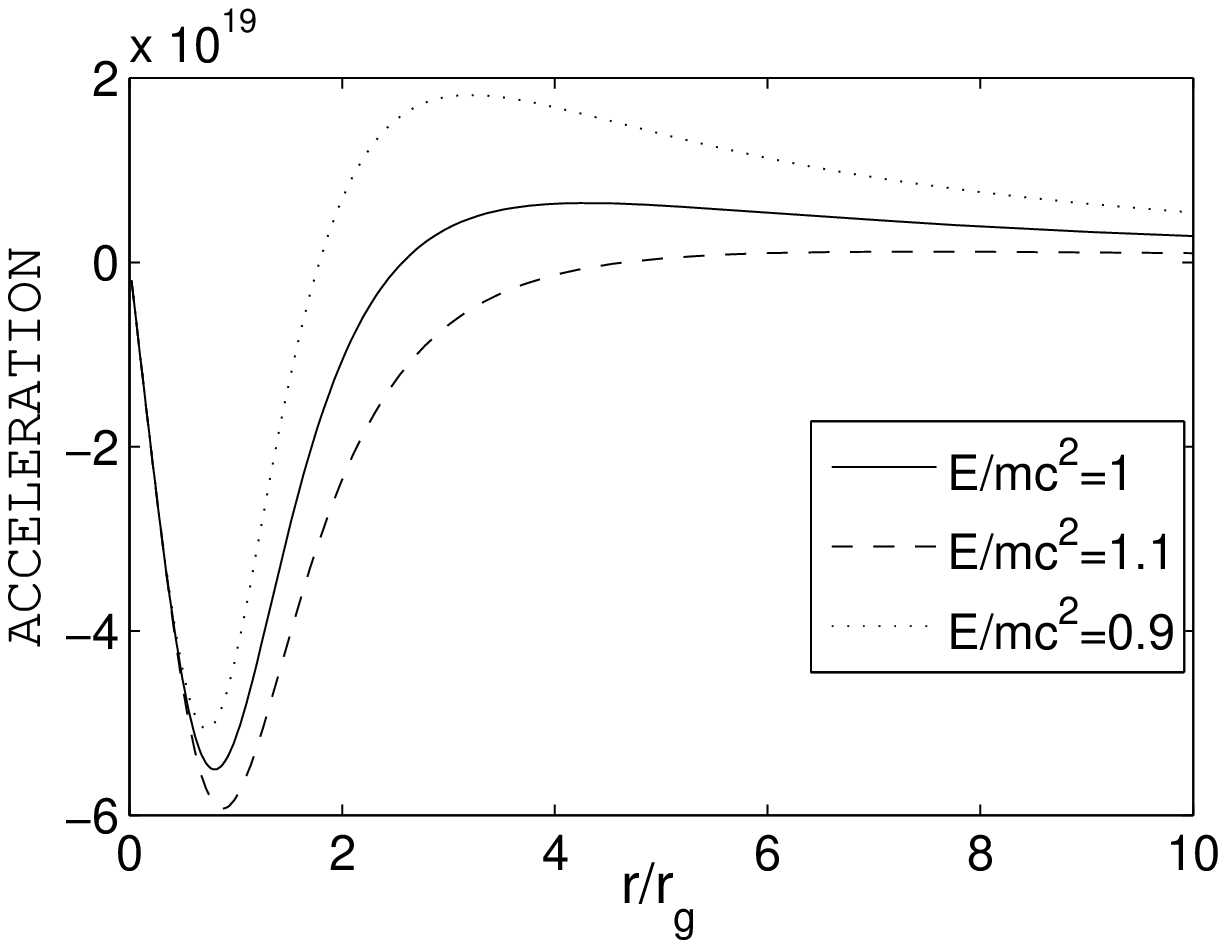} \caption{\label{fig:accelpart}The acceleration of a free falling test particle
(arbitrary units) near the attractive point mass}
\end{minipage}\hfill{}%
\begin{minipage}[t]{0.45\columnwidth}%
\includegraphics[width=4cm]{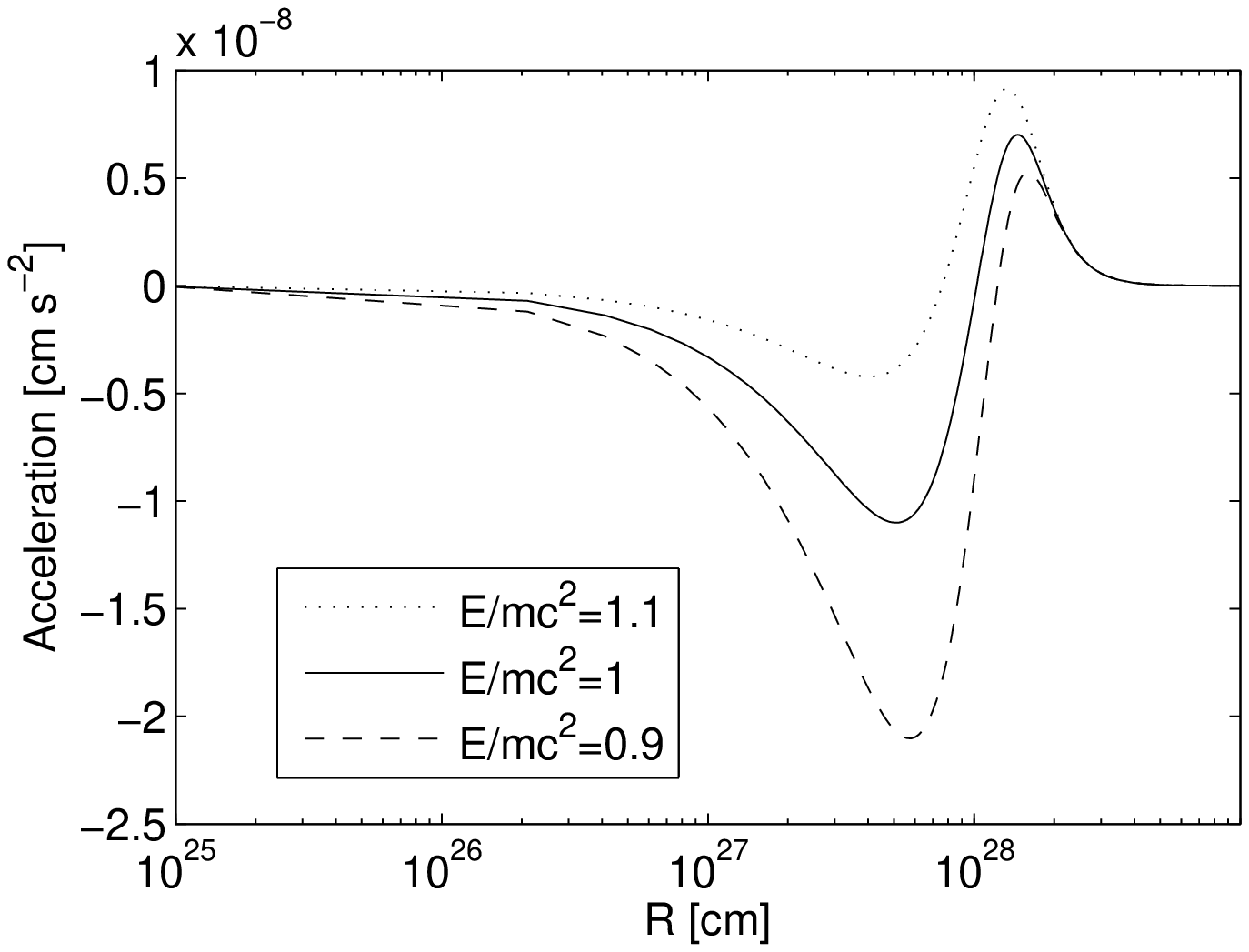} \caption{\label{fig:accelIniv}The acceleration of a test particle in the expanding
Universe vs. the distance from the observer.}
\end{minipage}
\end{figure}

Fig. \ref{fig:accelpart} shows a radial acceleration $w$ of a free
test particle as the function of the distance $\bar{r}=r/r_{g}$ from
the central dot mass $M$. This magnitude is $w=v\: dv/dr$ , where
$v$ is the radial velocity. The radial acceleration $w$ (or the
force $F=mw$ , acting on a test mass $m)$ is a finite on the all
interval ($0\div\infty$). If $r\gg r_{g}$, the acceleration $w=w(r)$
is the same as in the Newtonian mechanics. However, near $r_{g}$
the acceleration change the sign and eventually, inside the Schwarzschild
radius $r_{g}$ , the acceleration tend to zero when the distance
tends to zero.

Because of the peculiarity of the gravitational force $F=mw$ some
peculiar supermassive objects (up to $10^{10}M_{\odot}$ and more
than that) without the event horizon can exist with the radius less
than the Schwarzschild radius. Such objects are candidates to supermassive
objects in galactic centers \cite{Verozub2006,Verozub2008b}.

Consider now of an observer in the Minkowski space-time in the center
of a homogeneous dust-like sphere with the density of the order of
the Universe density ($10^{-27}\div10^{-28}g/cm^{3}$).\foreignlanguage{english}{
For him the radial acceleration of a test particle at the distance
$R$ at any moment depends only on the mass $M$ inside of the sphere
of the radius R. This acceleration is given by the external (vacuum)
solution of the eqs. (\ref{myeqs}) at $r=R.$}

The radial velocity at the surface of homogeneous sphere as the function
of its radius is given by (\ref{eq:rprime})

\begin{equation}
v(R)^{2}=c^{2}C(R)^{2}\left(1-\kappa^{2}C(R)/\bar{E}^{2}\right).\label{eq:velocity}
\end{equation}

Here $v(R)$ and $C(R)$ are given by eqs. (\ref{eq:ABC}), where
$r$ is replaced by $R$, $M=(4/3)\pi\rho\, R^{3}$ is the matter
mass inside of the sphere of the radius $R$, $\rho$ is the matter
density, and $r_{g}=(8/3)\pi c^{-2}G\rho R^{3}$ is Schwarzshild's
radius of the matter inside of the sphere.

Now add to the sphere outside some spherical shell of a mass $M_{1}$
and radius $R_{1}>R$. The mass of this shell does not change the
value of $w(R)$. The velocity and acceleration at $r=R_{1}$ is defined
at any moment by the full mass inside the external sphere and by the
radius $R_{1}$. Continuing to do the same one can find the dependence
of the radial acceleration on $r$ around the observer for very large
distances.

The mass of the homogeneous sphere is proportional to first degree
of density, and to third degree of the sphere radius. The observable
radius of the Universe is of order of $10^{28}cm$ that is of the
order of the Schwarzschild radius or less than that. Therefore, the
acceleration (or the force $mw(r)$, acting on a test body of mass
$m$) decreases, and at $r\rightarrow\infty$ tends to zero. At that
the functions $A(r),\, B(r),\, C(r)$ tend to their values in the
Minkowski space. Thus, space-time in PRFs of gravitational field of
a very large space distribution of a homogeneous mass is an Euclidean
on infinity, i.e is in full compliance with used conditions (\ref{limitConditions}).

Fig. 2 shows the radial acceleration of a test particle as the function
of the radius $R$ of a homogeneous sphere with real cosmological
parameters.

Two conclusions can be made from this figure.

1. If the density $\rho=6\cdot10^{-30}g\, cm^{-3}$, the Schwarzschild
radius $r_{g}$ becomes more then $R$ at $R>1.5\cdot10^{28}cm$.
Before this, at $R=6\cdot10^{27}cm$, the relative acceleration change
the sign, the same as for the particle free falling to a dot mass.
If $R>6\cdot10^{27}cm$, the acceleration is positive. Hence, for
sufficiently large distance $R$ the gravitational force gives rise
an acceleration of galaxies. It must be emphasized that all magnitudes
has sense also at the distances less than the Schwarzschild radius
of the considered masses.

2. The gravitational force, affecting the particles, tends to zero
when $R$ tends to infinity. The reason of the fact is that the ratio
$R/r_{g}$ tends to zero when $R$ tends to infinity. Consequently,
the gravitational influence on galaxies at large distance $R$ cased
mainly by the matter insider of the sphere of the radius $R$.

\section{The Hubble diagram\label{sec:4}}

A magnitude which is related with observations in the expanding Universe
is the relative velocity of a distant galaxy with respect to the observer.
The radial velocity $v$ is given by (\ref{eq:velocity}).

Proceeding from this result we will find Hubble diagram, following
mainly the method being used in \cite{Zeldovich}.

Let $\nu_{0}$ be a local frequency of light in the co-moving reference
frame of a moving source at the distance $R$ from an observer, $\nu_{l}$
be this frequency in a local inertial frame, and $\nu$ be the frequency
as measured by the observer in the center of the sphere with the radius
$R$.

The redshift $z=(\nu-\nu_{0})/\nu$ is caused by both the Doppler-effect
and gravitational field.

The Doppler-effect is a consequence of a difference between the local
frequency of the source in inertial and co-moving reference frame,
and it is given by \cite{Landau} 
\begin{equation}
\nu_{l}=\nu_{0}\left[(1-v/c)(1+v/c)\right]^{1/2}.\label{doppler}
\end{equation}

The gravitational redshift is caused by the matter inside of the sphere
of the radius $R$. It is a consequence of the energy conservation
for a photon. According to the equations of the motion of a test particle
(\ref{eq:rprime}) the rest energy of a particle in gravitation field
is given by 
\begin{equation}
E=mc^{2}\sqrt{C}.\label{eq:Egrav}
\end{equation}
Therefore, the difference in two local level $E_{1}$ and $E_{2}$
of an atom the energy in the field is $\Delta E=(E_{2}-E_{1})\sqrt{C}$,
so that %the local frequency $\nu_{0}$ at the distance $R$ from an
%observer are related with the observed frequency $\nu$ by the equality
\begin{equation}
\nu=\nu_{l}\sqrt{C},\label{redsiftGrav}
\end{equation}
where we take into account that for the observer location $C=1$.
It follows from (\ref{doppler}) and (\ref{eq:Egrav}) that the relationship
between the frequency $\nu$ , as measured by the observer, and the
proper frequency $\nu_{0}$ of the moving source in the gravitational
field takes the form

\begin{equation}
\frac{\nu}{\nu_{0}}=\sqrt{C\frac{1-v/c}{1+v/c}}\label{eq:nu-nu0}
\end{equation}

This equation yields the quantity $z=\nu_{0}/\nu-1$ as a function
of $R$. By solving this equation numerically we obtain the dependence
$R=R(z)$ of the measured distance $R$ as a function of the redshift.Therefore,
the distance modulus for a remote galaxy is given by

\begin{equation}
\mu=5\, log_{10}[R(z)\,(z+1)]-5\label{eq:mu}
\end{equation}
where $R(1+z)$ is a bolometric distance (in $pc$) to the object.

If eq.(\ref{eq:velocity}) have to give a correct radial velocity
of distant of the galaxy in the expansive Universe, it have to lead
to the classic Hubble law at small distances of $R$. At this condition
the Schwarzschild radius $r_{g}=(8/3)\pi G\rho R^{3}$ of the matter
inside of the sphere is very small compared with $R$. For this reason
$f\approx r$, and $C=1-r_{g}/r$. Therefore, at $\overline{E}=1$,
we obtain from (\ref{eq:velocity}) that 
\[
v=HR,
\]
where 
\[
H=\sqrt{(8/3)\pi G\rho}.
\]

If $\overline{E}\neq1$, then equation (\ref{eq:velocity}) does not
lead to the Hubble law since $v$ does not tend to zero when $R\rightarrow0$.
For this reason we set $\overline{E}=1$ and look for the value of
the density for which a good accordance with observation data can
be obtained.

\begin{figure}[H]
\begin{minipage}[t]{0.45\columnwidth}%
\includegraphics[width=4cm]{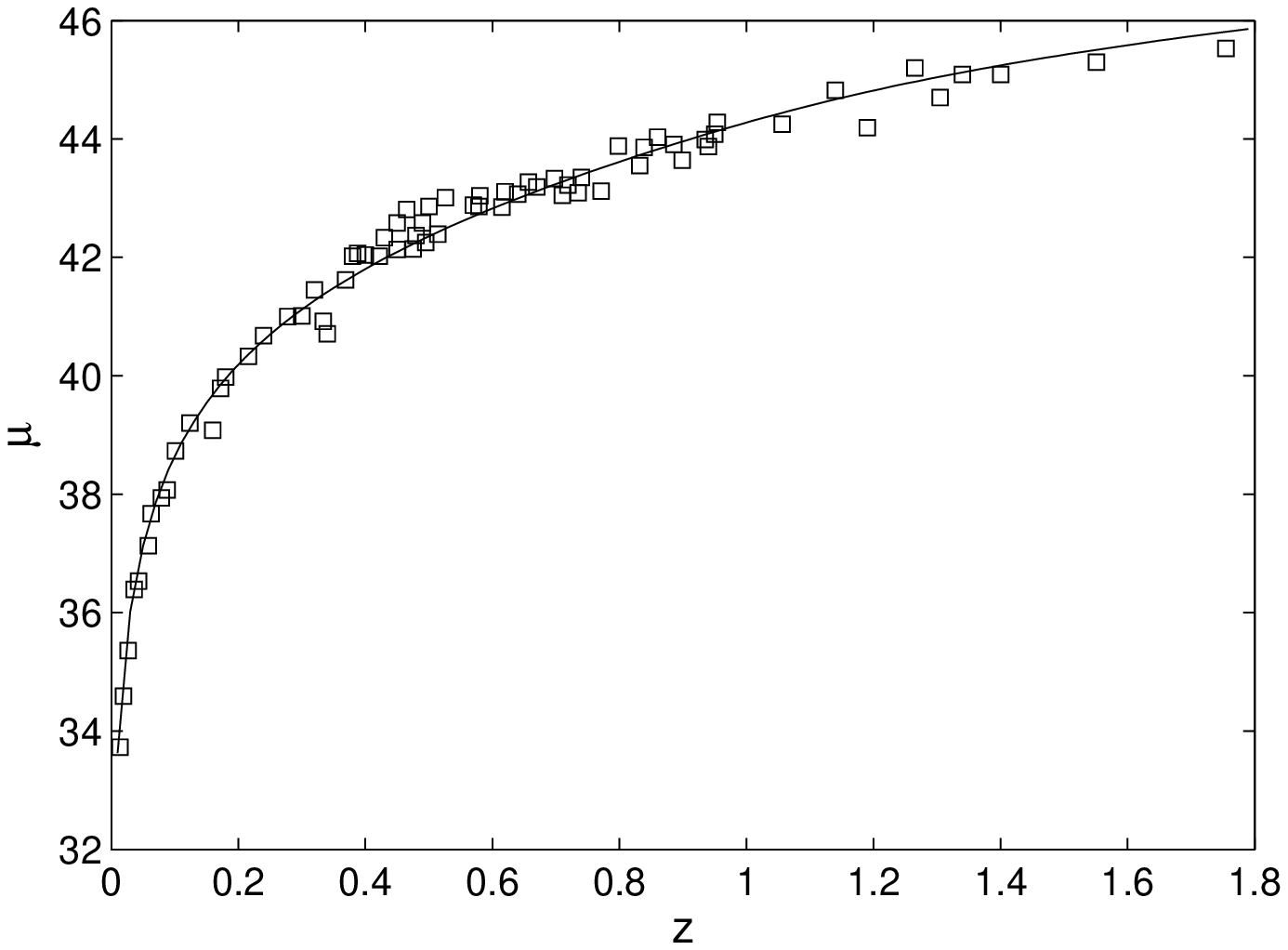} \caption{\label{fig:muofzold}The distance modulus $\mu$ vs. the redshift
$z$ for the density $\rho=6\cdot10^{-30}g\, cm^{-3}.$ The small
squares denote the observation data according to \cite{Riess}.}
\end{minipage}\hfill{}%
\begin{minipage}[t]{0.45\columnwidth}%
\includegraphics[width=4cm]{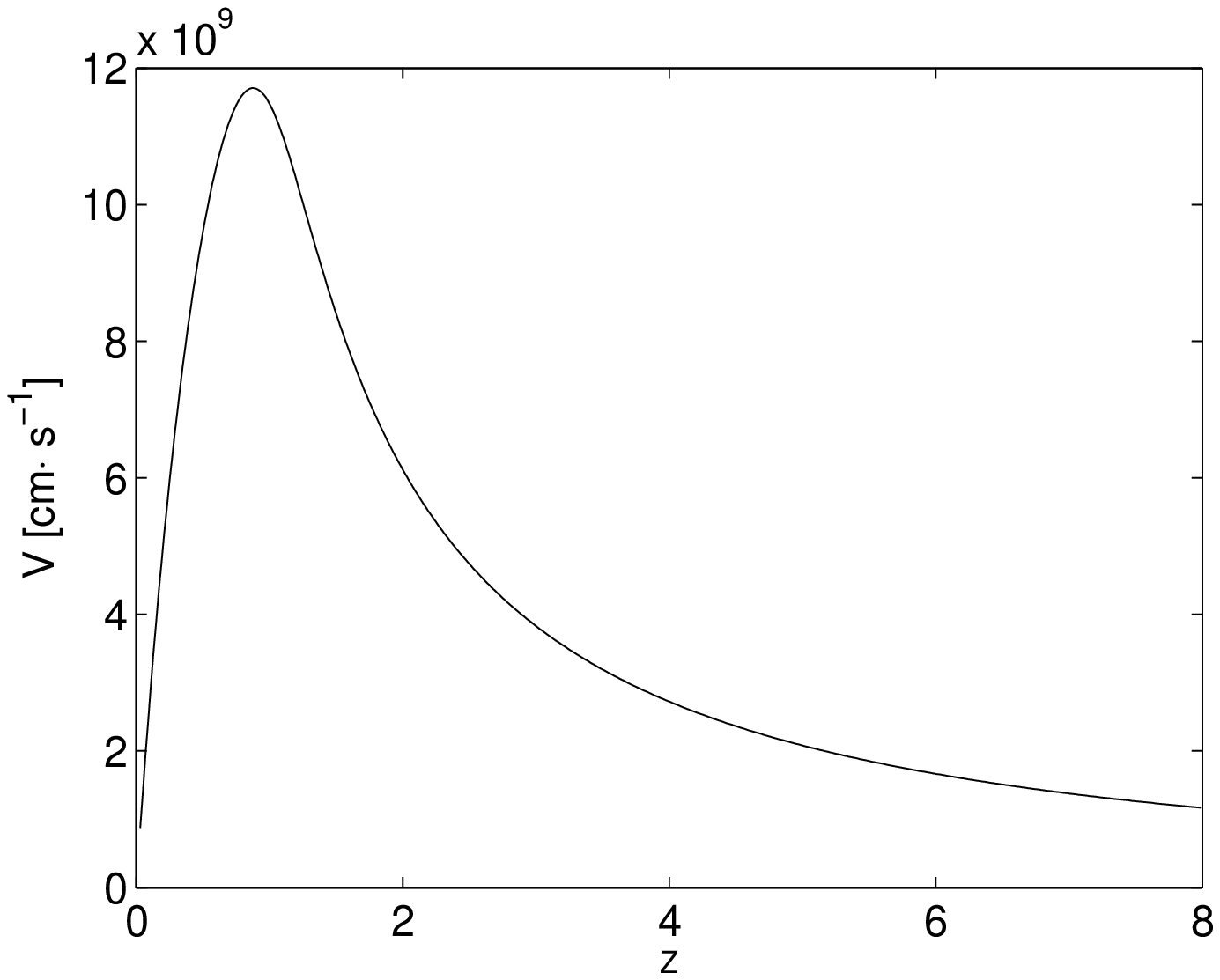} \caption{\label{fig:fvofz}The radial velocity vs. redshift $z$for the density
$\rho=6\cdot10^{-30}g\, cm^{-3}$.}
\end{minipage}
\end{figure}

%\selectlanguage{english}%
The fig.\ref{fig:muofzold} shows the Hubble diagram up to $z=1.8$
obtained by eq.(\ref{eq:mu}) compared with observations data obtained
by supernova Ia \cite{Riess}. A good agreement between theory and
observation is obvious.

%\selectlanguage{american}%
Figure \ref{fig:fvofz} shows the dependence of $v$ on $z$ up to
$z=8$.

\begin{figure}[H]
\begin{minipage}[t]{0.45\columnwidth}%
\includegraphics[width=4cm]{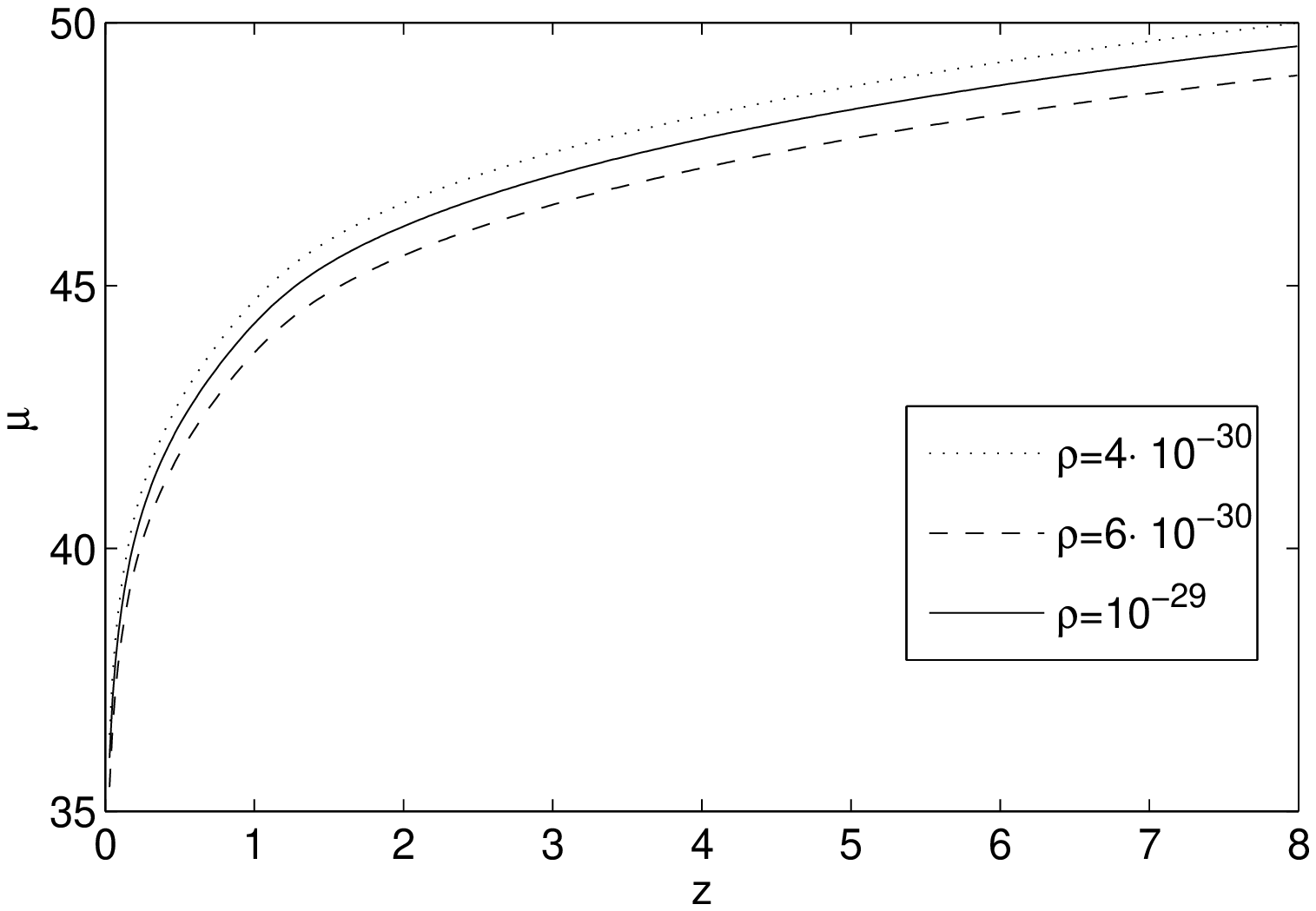} \caption{\label{fig:muofznew}The Hubble diagram according to (\ref{eq:mu})
up to $z=8$ for several magnitudes of the matter density $\rho$. }
\end{minipage}\hfill{}%
\begin{minipage}[t]{0.45\columnwidth}%
\includegraphics[width=3.7cm]{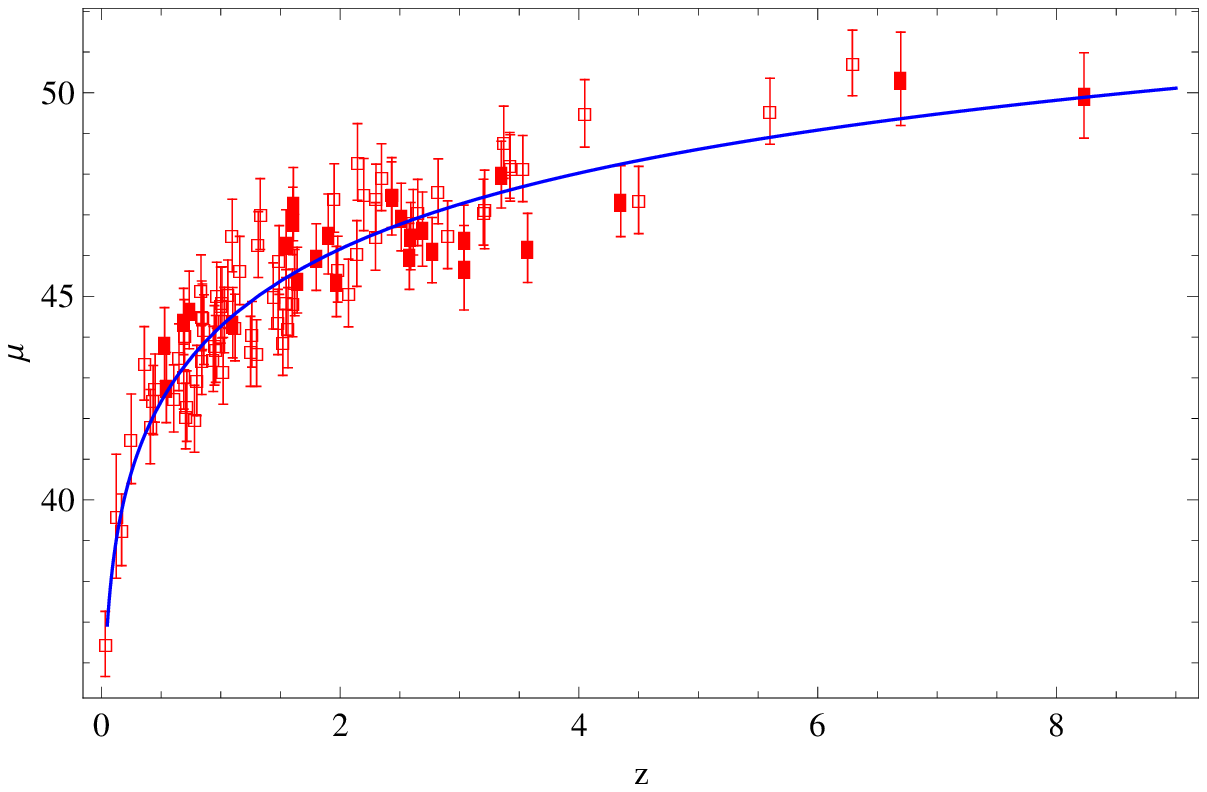} \caption{\label{fig:muofzdemianski}The Hubble diagram \cite{Demiansky} obtained
by using GRBs data.}
\end{minipage}
\end{figure}

Figure \ref{fig:muofznew} shows the distance modulus obtained by
the above method as a function of $z$ up to $z=8$. Figure \ref{fig:muofzdemianski}
reproduces the typical dependence of the distance modulus on $z$
up to $z=8$ obtained with an analysis of gamma-ray bursts \cite{Cardone,Demiansky,Schaefer}.

%\begin{minipage}[t]{0.45\columnwidth}%

A comparing the last two plots show good agreement between the theory
and observation.

%\selectlanguage{english}%

%\selectlanguage{american}%

\section{Conclusion\label{sec:5}}

In paper \cite{Verozub2006} was shown that a specific properties
of gravity in the vicinity of the Schwarzschild radius does not contradict
the observational data.

In this paper we have shown that just such properties of the gravitational
force allow us to understand the well known peculiarity of the Hubble
diagram at large redshifts up to $z=8$ that indicates an acceleration
of the Universe expansion.

It should be noted that the model actually uses only one fitting parameter
- the density.

This result shows that high redshift can be an important instrument
for testing theory of gravitation.

\section{Supplement\label{sec:6}}

The most interesting examples of non-uniqueness of the metric of space-time
in a given coordinate system due to the existence of the geodesic
equivalence are FRW and Schwarzschild metrics. Here is a proof of
this fact, based on papers\cite{Mikes,VerozubEprint}, which shows
that the metrics obtained from a given by some continuous one-parameter
transformations have common geodesies .

\subsection*{FRW Metric\label{sub:6.1}}

Consider the line elements of a Riemannian space-time $V$:

\begin{equation}
ds^{2}=b(t)\, dt^{2}+a(t)\,\sigma_{ik}(x^{1},x^{2},x^{3})\, dx^{1}dx^{k}.\label{eq:ds}
\end{equation}
Geodesics of such metric are the same as the ones of the space-time
$\overline{V}$ with the line element 
\begin{equation}
\overline{ds}^{2}=B(t)\, dt^{2}+A(t)\,\sigma_{ik}(x^{1},x^{2},x^{3})\, dx^{k}dx^{k}.\label{eq:dsprime}
\end{equation}

where

\begin{equation}
B(t)=\frac{b(t)}{[1+q\, a(t)]^{2}},\label{eq:A}
\end{equation}

\begin{equation}
A(t)=\frac{a(t)}{1+q\, a(t)},\label{eq:B}
\end{equation}
and $q$ is an arbitrary constant . The proof of this important fact
is follows.

Contracting (\ref{gaugeTrans}) with respect to $\alpha$ and $\beta$,
we obtain $\overline{\Gamma}_{\beta\gamma}^{\beta}=\Gamma_{\beta\gamma}^{\beta}+(n+1)\,\psi_{\beta}.$
Consequently, 
\begin{equation}
\psi_{\beta}=\frac{1}{2(n+1)}\,\frac{\partial}{\partial x^{\beta}}\ln\left|\cfrac{\det\overline{g}}{\det g}\right|,\label{eq:psi}
\end{equation}
which shows that in the case under consideration only $0$-component
of $\psi_{\alpha}$is other than zero.

The useful for us components of the Christoffel symbols of $V$ are
given by:

\begin{multline*}
\overline{\Gamma}_{00}^{0}=b'(t)/2\, b,\quad\overline{\Gamma}_{11}^{0}=a'(t)/2b(t),\\
\overline{\Gamma}_{10}^{1}=a'(t)/2\: a(t),
\end{multline*}
and the same components of $\overline{V}$ are: 
\begin{multline}
\overline{\Gamma}_{00}^{0}=B'(t)/2\, B,\quad\overline{\Gamma}_{11}^{0}=-A'(t)/2B(t),\\
\overline{\Gamma}_{10}^{1}=A'(t)/2\: A(t)
\end{multline}
Then eqs. (\ref{gaugeTrans}) gives the following equations

\begin{equation}
\frac{A'(t)}{A(t)}-\frac{a'(t)}{a(t)}=2\,\psi_{0},\label{eq:dif1}
\end{equation}

\begin{equation}
\frac{B'(t)}{B(t)}-\frac{b'(t)}{b(t)}=4\,\psi_{0},\label{eq:dif2}
\end{equation}

\begin{equation}
\frac{A'(t)}{B(t)}-\frac{a'(t)}{b(t)}=0.\label{eq:dif3}
\end{equation}
So, $A/a=\exp(2\int\psi(t)_{0}dt)$, $B/b=\exp(4\int\psi(t)_{0}dt)$
where the integration constants are equal to $1$ because at $\psi(t)_{0}=0$
the functions $A(t)=a(t)$ and $B(t)=b(t)$. Consequently, $B(t)/b(t)=(A(t)/a(t))^{2},$and
with (\ref{eq:dif3}) we obtain the differential equations 
\begin{equation}
A'(x)-A(t)^{2}\frac{a'(t)}{a(t)}=0,
\end{equation}
which gives (\ref{eq:B}). Now from previous equation we obtain the
function $B(t)$ in the form (\ref{eq:A}).

On the contrary, if in eq. (\ref{gaugeTrans}) to set $\psi_{i}=0$
for i=1,2,3, and $\psi_{0}=-\nicefrac{1}{2\,}\partial$$\ln(1+qb(t))/\partial t$,
then eqs. \eqref{eq:dif1}, \eqref{eq:dif2}, and \eqref{eq:dif3}
are satisfied. Thus, with this choice of the co-vector field $\psi(x)_{\alpha}$,
the line element \eqref{eq:ds} at $b=-1$ is equivalent to \eqref{eq:dsprime}.
In other words, the both line elements have the same (no-parameterized)
equations of motion of test particles.

\subsection*{Schwarzschild Metric\label{sub:6.2}}

As another example, we show here that a static centrally symmetric
metric

\begin{eqnarray}
ds^{2} & = & b(r)dr^{2}+r^{2}(d\theta^{2}+\sin^{2}\theta d\phi^{2})-a(r)dt^{2},\label{eq:ds-1}
\end{eqnarray}
(in particular, Shirtsleeve metric) in a given coordinate system is
not unique. Namely, in any given coordinate system it has common geodesic
lines with a metric of the form

\begin{equation}
\overline{ds}^{2}=B(r)\, dr^{2}+F(r)^{2}(d\theta^{2}+\sin^{2}\theta\, d\phi^{2})-A(r)\, dt^{2},\label{eq:dsprime-1}
\end{equation}
where $A(x),\: B(x)$ and $F(x)$ are functions of $x^{\alpha}$,
depending on a continuous parameter.

The Christoffel symbols for (\ref{eq:ds-1}) is given by

\begin{multline}
\Gamma_{rr}^{r}=\frac{1}{2}\frac{b'(r)}{b(r)},\;\Gamma_{r\theta}^{\theta}=\frac{1}{r},\;\Gamma_{r\phi}^{\phi}=\frac{1}{r},\;\Gamma_{rt}^{t}=\frac{1}{2}\frac{a'(r)}{a(r)},\\
\Gamma_{\theta\theta}^{r}=-\frac{r}{b(r)},\label{eq:GammaV1}
\end{multline}

\begin{multline}
\Gamma_{\theta\phi}^{\phi}=\frac{\cos\theta}{\sin\theta},\;\Gamma_{\phi\phi}^{r}=-\frac{r\sin^{2}\theta}{b(r)},\;\Gamma_{\phi\phi}^{\theta}=-\sin\theta\cos\theta,\\
\Gamma_{tt}^{r}=\frac{1}{2}\frac{a'(r)}{b(r)}.\label{eq:GammaV2}
\end{multline}
The Christoffel symbols for \ref{eq:dsprime-1} are:

\begin{multline}
\overline{\Gamma}_{rr}^{r}=\frac{1}{2}\frac{B'(r)}{B(r)},\;\overline{\Gamma}_{r\theta}^{\theta}=\frac{F'(r)}{F(r)},\;\overline{\Gamma}_{r\phi}^{\phi}=\frac{F'(r)}{F(r)},\\
\overline{\Gamma}_{rt}^{t}=\frac{1}{2}\frac{A'(r)}{A(r)},\overline{\Gamma}_{\theta\theta}^{r}=-\frac{F(r)\, F'(r)}{B(r)},\label{eq:GammaVprime1}
\end{multline}

\begin{multline}
\overline{\Gamma}_{\theta\phi}^{\phi}=\frac{\cos\theta}{\sin\theta},\;\overline{\Gamma}_{\phi\phi}^{r}=-\frac{F(r)\, F'(r)\sin ApJ^{2}\theta}{B(r)},\\
\overline{\Gamma}_{\phi\phi}^{\theta}=-\sin\theta\cos\theta,\overline{\Gamma}_{tt}^{r}=\frac{1}{2}\frac{A'(r)}{B(r)},\label{eq:GammaVprime2}
\end{multline}
where a prime here and later denotes a derivative with respect to
$r$.

In view of this, Levi-Chevita equations (\ref{gaugeTrans}) yields:
\begin{multline}
\frac{B'(r)}{B(r)}-\frac{b'(r)}{b(r)}=4\psi_{r}(r),\; F(r)F'(r)b(r)-rB(r)=0,\\
A'(r)b(r)-a'(r)B(r)=0,\label{eq:eqs1}
\end{multline}

\begin{multline}
\frac{F'(r)}{F(r)}-\frac{1}{r}=\psi_{r}(r),\;\frac{A'(r)}{A(r)}-\frac{a'(r)}{a(r)}=2\psi_{r}(r),\\
\psi_{\theta}(r)=\psi_{\phi}(r)=\psi_{t}(r)=0\label{eq:eqs2}
\end{multline}

According to (\ref{eq:psi}) the function $\psi_{r}(r)$ . can be
written as 
\[
\psi_{r}=\partial\ln\chi/\partial r,
\]
where 
\[
\chi(r)=\left(\frac{\bar{g}}{g}\right)^{1/2(n+1)}.
\]
Consequently,

\begin{center}
$\begin{array}{ccc}
B=b\chi^{4}; & A=a\chi^{2}; & F=\chi;\\
A'=a'\chi^{4}; & \left(F^{2}\right)^{'}=2r\chi^{4};
\end{array}$ 
\par\end{center}

Formulas for the function $F(r)$ are compatible only if the functions
$\chi(r)$ are the solution of the differential equations

\[
r\chi'(r)+\chi(r)-\chi(r)^{3}=0
\]
which yields 
\[
\chi(r)=(1+kr^{2})^{1/2},
\]
where $k$ is an arbitrary constant.

As a result, formulas which express the $A(r),B(r)$, and $F(r)$
by $a(r)$ and $b(r)$ are given by

\begin{multline}
A(r)=\frac{a(r)}{1+kr^{2}},\; B(r)=\frac{b(r)}{\left(1+kr^{2}\right)^{2}},\\
F(r)=\frac{r}{(1+kr^{2})^{1/2}}.\label{eq:AandB}
\end{multline}

%\selectlanguage{english}%

\section*{}
\end{document}